# Study of Vision Transformers for Covid-19 Detection from Chest X-rays

Sandeep Angara
sandeepangara@gmail.com

Sharath Thirunagaru
sharath.tds@gmail.com

*Abstract*—The COVID-19 pandemic has led to a global health crisis, highlighting the need for rapid and accurate virus detection. This research paper examines transfer learning with vision transformers for COVID-19 detection, known for its excellent performance in image recognition tasks. We leverage the capability of Vision Transformers to capture global context and learn complex patterns from chest X-ray images. In this work, we explored the recent state-of-art transformer models to detect Covid-19 using CXR images such as vision transformer (ViT), Swin-transformer, Max vision transformer (MViT), and Pyramid Vision transformer (PVT). Through the utilization of transfer learning with IMAGENET weights, the models achieved an impressive accuracy range of 98.75% to 99.5%. Our experiments demonstrate that Vision Transformers achieve state-of-the-art performance in COVID-19 detection, outperforming traditional methods and even Convolutional Neural Networks (CNNs). The results highlight the potential of Vision Transformers as a powerful tool for COVID-19 detection, with implications for improving the efficiency and accuracy of screening and diagnosis in clinical settings.

Keywords— Vision transformers, Swin-transformers, Covid-19, Chest X-rays

## I. INTRODUCTION

Covid-19, also known as the coronavirus, is an infectious disease caused by severe acute respiratory syndrome coronavirus 2 (SARS-CoV-2) [1]. First identified in Wuhan, China, in December 2019, the disease quickly spread to other countries, leading the World Health Organization (WHO) to declare it a pandemic in March 2020. Since its emergence in late 2019, it has profoundly impacted the world. The virus is primarily spread through respiratory droplets when an infected person talks, coughs, or sneezes and can be contracted by anyone who comes into close contact with an infected individual or surface—people infected with covid experience mild to moderate respiratory illness [2][3]. The disease has affected millions of people worldwide, leading to significant illness and loss of life. The pandemic has put a strain on healthcare systems, caused economic disruptions, and changed daily life in many ways. Covid-19 exposed the vulnerability of health workers and professionals worldwide, indicating a clear need for automatic diagnosis tools for detection. Since the virus affects the lungs, Chest X-rays can help diagnose and manage COVID-19. Any method for automatic, reliable, and accurate screening of COVID-19 infection would be beneficial for rapid detection and reducing medical or healthcare professional exposure to the virus.

Detecting COVID-19 has been essential in containing its spread and finding people who have been infected. Different techniques have been created to identify the presence of SARS-CoV-2, the virus accountable for COVID-19. Three commonly used techniques to detect Covid-19 are polymerase chain reaction (PCR), antigen testing, and antibody testing [4][5]. PCR is currently considered the gold standard for COVID-19 detection. This highly sensitive and specific molecular diagnostic technique detects the presence of the genetic material of the SARS-CoV-2 virus, specifically its RNA, in a patient's respiratory sample. PCR tests offer high sensitivity and specificity, allowing for accurate diagnosis even in the early stages of infection. However, they require specialized laboratory equipment and trained personnel, which can result in longer turnaround times and higher costs.

Antigen tests detect viral proteins (antigens) in respiratory samples and are faster and cheaper than PCR tests. They can provide results within minutes, identifying infected individuals immediately. However, their sensitivity is lower than PCR tests, especially in the early stages of infection when the viral load is low. False negatives can occur, so PCR tests may sometimes be needed for confirmation.

Serological tests, also called antibody tests, can identify the presence of antibodies created by the immune system due to SARS-CoV-2 infection. These tests are conducted on blood samples and can show previous infection or an immune reaction to vaccination. Although not as useful for detecting active infections, antibody tests may not identify antibodies during the initial stages of the illness. Antibody tests are useful for determining past infections but are less effective for diagnosing active cases. The testing method chosen should depend on the specific needs, resources, and objectives of the testing program, which may include screening populations, diagnosing acute infections, or assessing immunity within communities.

Additionally, covid can be detected using chest X-rays and computed tomography (CT) [6]. Chest X-rays provide a quick way to assess lung damage caused by the virus. Chest X-rays are more helpful in monitoring the progression of the disease, evaluating the severity of lung damage, and identifying potential complications [7]. CT scans of the chest can also show lung damage caused by COVID-19 and provide more detailed images of the lungs than chest X-rays [8]. However, CT scans are not routinely used in COVID-19 diagnosis due to their higher cost, longer imaging time, and increased exposure to radiation.

Since the start of Covid-19, researchers have explored deep learning in detecting CXR or CT images [9][10][11][12][13]. Automating the diagnosis process helps to save time and effort for radiologists, which is a long and error-prone process. Deep learning algorithms have shown promise in assisting with the detection of COVID-19 using chest X-rays. Several studies have reported using deep learning algorithms to analyze chest X-rays to detect COVID-19. These algorithms have been trained on large datasets of chest X-rays from COVID-19 positive and negative patients to learn patterns and features that can distinguish between the two. The performance of these algorithms in COVID-19 detection using chest X-rays has been reported to be comparable to or even better than that of human radiologists. One of the benefits of using deep learning algorithms in COVID-19 detection is their ability to quickly analyze large volumes of medical images, which can be particularly useful in areas with a shortage of radiologists or during a pandemic like COVID-19. Furthermore, deep learning algorithms can be trained to identify other disease features, such as the severity and progression of lung damage.

## II. RELATED WORKS

Deep learning has shown great potential in medical imaging, which can assist in analyzing and interpreting medical images, such as X-rays [14][15][16], CT scans[17][18], MRI scans [19][20][21][22], ultrasound images[23][24] etc. Medical imaging technology has made significant advances with the help of deep learning. The potential benefits of using this technology in healthcare are immense, and ongoing research shows promising results. Researchers have been actively exploring the use of Chest X-rays for COVID-19 detection due to the increased popularity of deep learning in medical imaging.

Wang et al. [25] proposed COVID-NET, a deep convolutional neural network design tailored for COVID-19 detection from chest X-rays, which was trained on three classes, i.e., Normal, Pneumonia, and COVID-19. COVID-Net is designed by heavy usage of lightweight residual projections extension design, enabling enhanced representational capacity with reduced computational capacity, which has an accuracy of 93.3%. Das et al. [26] used transfer learning to train the Xception network for COVID-19 classification. Experiments were conducted on 127 COVID-19, 500 Controls, and 500 Pneumonia samples collected from various public datasets, achieving an accuracy of 97%. Krishnan et al. [27] proposed a modified vision transformer architecture trained using transfer learning to detect covid from chest X-rays and achieved an accuracy of 97.61%. The experiments were conducted on COVIDx CXR-2 dataset, which consists of 19105 samples. Guefrechi et al. [28] finetuned three powerful networks VGG16[29], ResNet50 [30], and InceptionV3[31] on COVID-19 dataset constructed by collecting COVID-19 and normal chest X-ray images from different public databases. Out of all three architectures, VGG16 outperformed with an accuracy of 98.30%.

Islam et al. [10] developed a CNN-LSTM-based network to detect COVID-19; the network was trained on a dataset with three classes COVID-19, Pneumonia, and normal. The dataset consists of 3660 images in the training set and 915 test datasets. The proposed system achieved an accuracy of 99.4%, an AUC of 99.9%, a specificity of 99.2%, a sensitivity of 99.3%, and an F1-score of 98.9%. Pavlova et al. [32] proposed a lightweight network named COVID-NET CXR-2 designed based on the originally proposed COVID-Net [25]. The network possesses a mix of point-wise and depth-wise convolutions, which reduced the computational capacity by ~25% compared to COVID-Net. In addition, the new architecture achieved a test accuracy of 96.3% and an area under the curve (AUC) of 99.4%.

Guang et al. [33] explored self-supervised transfer learning to detect COVID-19 from chest X-rays. The authors also compared self-supervised techniques with transfer learning based on Resnet18, ResNet50, ResNet101, InceptionV3, and DenseNet121. Self-supervised method outperformed the transfer learning by a significant margin and achieved an accuracy of 95.3% on four class datasets. Junghoon et al. [34] developed a model for diagnosing COVID-19 using a self-supervised learning technique with a convolution attention module. In this work a U-shaped supervised learning technique with a convolutional neural network model combined with a convolutional block attention module (CBAM) using over 100,000 chest X-Ray images with structure similarity (SSIM) index to learn image representations extremely well. The final classifier is finetuned on the encoder weights learned by self-supervised learning. The average accuracy of the classifier is 98.6%.

Debaditya et al. [35] proposed a custom transformer model that effectively discriminates COVID-19 from normal chest X-rays with an accuracy of 98% and AUC score of 99%. A new self-supervised paradigm was proposed by Syed et al. [36] that involves learning a general representation from CXRs through a group-masked self-supervised framework. The pre-trained model can then be fine-tuned for tasks specific to domains like covid-19, pneumonia detection, and general health screening. The authors have used the VIT-S model for representation learning and achieved an accuracy of 98.25%.

Most previous studies use chest X-rays to detect COVID-19, highlighting the importance of image analysis as a trustworthy diagnostic tool for doctors. It is widely understood that a large amount of labeled data is required to train deep learning models effectively. Most studies explored COVID CXR-2 and other datasets, or curated datasets based on publicly available datasets. Very few studies explored the transformer-based models, which are the current state of art techniques outperforming the CNN counterparts.

In this paper, we performed a fine-grained study on transfer learning using transformer architectures. We explored vision transformer (ViT) [37], swin-transformer [38], multi-Axis vision transformer (MaxViT) [39], and pyramid vision transformer (PVT v2) [40]. We investigated the COVIDx CXR-3 dataset [40], the largest open-source benchmark dataset to date for chest X-ray images for computer-aided COVID-19 diagnostics. To the best of our knowledge, this is the first study to focus on examining multiple transformers architectures for detecting COVID-19 Limited research has been conducted on this topic thus far. We verify that the transformer models with transfer learning outperform the existing state of art techniques in COVID-19 detection from CXR images.

III. METHODOLOGY

*A. Datasets*

Our study used the COVIDx CXR-3 dataset [41], which contains a vast collection of chest X-rays from Covid-19 and pneumonia patients. This dataset, curated by expert radiologists and clinicians, is a benchmark for CXR images and includes images from patients across multiple institutions and countries. The COVIDx CXR-3 collection includes 30386 chest X-ray images taken from more than 17,026 patients across at least 51 countries; the CXR-3 dataset is one of the largest available. The x-ray dataset was split into a training set of 29986 images and a test set of 400 images. In the training set, there were 8,437 x-rays of normal cases, 5,555 x-rays of pneumonia cases, and 15994 x-rays of Covid-19 cases. The test set included 100 x-rays of normal cases, 100 x-rays of pneumonia cases, and 200 x-rays of Covid-19 cases.

*B. Architectures*

In this study, we explored transformer-based architectures in COVID recognition from X-rays. Transformers [42] are a deep learning architecture primarily used for natural language processing tasks. However, recent research has shown that transformers can also be effective in computer vision tasks, such as image recognition, object detection, and segmentation. Transformers use self-attention mechanisms to process input data, allowing the model to focus on relevant parts of the input data at different levels of abstraction. This has improved performance on image recognition tasks, particularly on datasets with large amounts of data and complex relationships between different image features. The success of transformers in computer vision has led to the development of new transformer-based architectures, such as the Vision Transformer (ViT) [37] and Swin Transformer [38], which have achieved state-of-the-art results on several benchmark image recognition and object detection datasets. The use of transformers in computer vision is a rapidly evolving area of medical imaging research [43] [44] [45]. In the experiments, the following modeified state-of-art transformer-based models have been used.

*1) Vision Transformer (ViT)*

The Vision Transformer (ViT) [37] is a paper that proposed a novel deep learning architecture for image recognition tasks. Traditional convolutional neural networks (CNNs) have been the dominant architecture in image recognition, but ViT introduced a new approach that uses self-attention mechanisms to process images. Vision Transformers consist of a series of transformer blocks with an additional patch embedding layer. The patch embedding layer splits the input image into fixed-size patches and maps it into a high-dimensional vector representation; each vector representation is coupled with a trainable positional embedding. The patches are linearly embedded with an additional trainable CLS token for classification tasks. Each transformer block includes a multi-head self-attention layer, feed-forward layer, and norm layer. The multi-head self-attention layer computes attention between a pixel with all other pixels, while many attention heads help to learn local and global dependencies in an image. A normalization layer is applied before each multi-head self-attention module, and a feed-forward layer. In this experiment, the model is presented with fixed-size patches of 16x16 embedded in a linear sequence. The final output layer is modified to two classes to recognize Normal and Covid CXRs, as shown in Fig. 1.

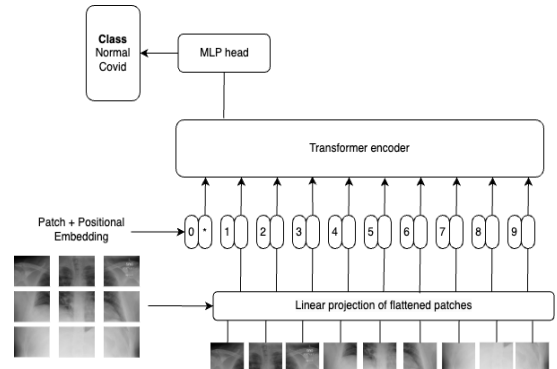

Fig 1. Vision Transformer Architecture

*2) Swin-transformer*

Swin-Transformer [38] is a recent deep learning architecture proposed in a research paper that has achieved state-of-the-art results in image recognition tasks. Swin transformer addresses two key issues faced by ViT i.e., hierarchical feature maps and shifted window attention that allows for better scalability and efficiency. The hierarchical architecture has the flexibility to model at various scales and has linear computational complexity with respect to image size. The model is organized into a series of stages, as shown in Figure 2; within each stage, the model uses self-attention to process image patches and extract features. The input image is split into a fixed patch size of 4x4 and passed to a linear embedding layer to project into tokens with an arbitrary dimension C. To produce hierarchical features, the number of tokens is reduced by patch merging as the network goes deeper. Swin-transformer has different variants, and we have used Swin-B for the experiment where C=128 and the output layer of the network is modified to classify the input image as Normal and Covid (Fig. 2).

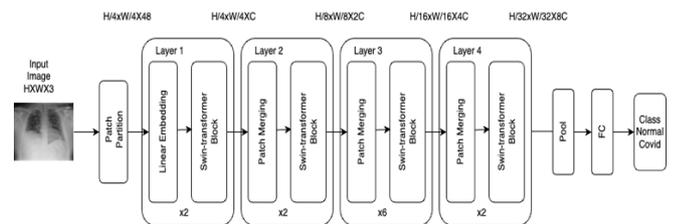

Fig 2. A Swin-Transformer architecture adapted from [38]

*3) Multi-Axis Vision Transformer (MaxViT)*

MaxViT [46] introduces a multi-axis self-attention (Max-SA) mechanism to capture local and global spatial interactions in a block. Max-SA decomposes the fully dense attention mechanisms into two sparse forms – window attention and grid attention, which provide linear complexity. Compared to (shifted) window/local attention, Max-SA can enhance model capacity by offering a global receptive field. MaxViT is designed hierarchically by stacking Max-SA blocks and convolutions. The multi-axis attention module consists of both blocked local and dilated global attention, enabling global perception at a linear complexity level. The architecture of MaxViT is built by cascading alternative layers of Max-SA with MBConv layers, as shown in Figures 3A & 3B. Each Max-SA module has block attention and grid attention, the former computes the local attention, and the latter computes the global attention. To increase the generalization, an MBConv block is added prior to block and grid attention blocks. MaxViT base model has been used for the training by modifying the final layer of the architecture to classify the input CXR image as Normal or Covid (Fig. 3).

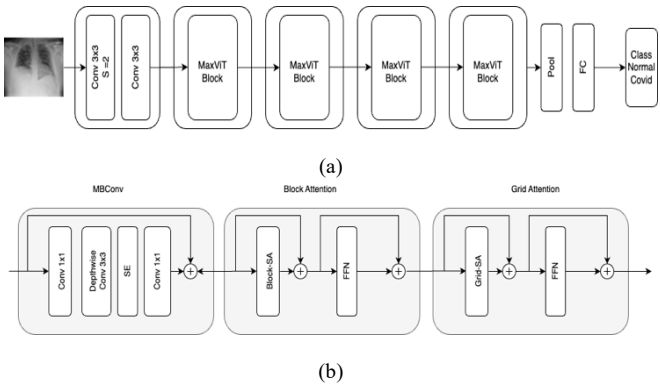

Fig. 3. (a) Multi-Axis Vision Transformer (MaxViT) architecture (adapted from [46]) (b) MaxViT block

*4) Pyramid Vision Transformer (PVT) v2*

Pyramid vision transformer [47] is designed to overcome the drawbacks of Vision transformer (ViT), as the output from ViT is not ideal for segmentation tasks due to its low resolution and single-scale feature map. PVT v2 [40] is built on a baseline model pyramid vision transformer with several novel improvements, which include a linear complexity attention layer, overlapping patch embedding, and convolution feed-forward network. With these modifications, the model has achieved state-of-the-art results on several benchmark image recognition datasets, including ImageNet, and has shown to be effective on other computer vision tasks, such as object detection and segmentation.

*C. Performance Criteria*

To thoroughly evaluate the effectiveness of our trained model, we utilize a range of metrics that have been widely accepted as reliable assessments in the medical field. While accuracy serves as the primary measure for evaluating classification performance, we also employ specific metrics such as sensitivity, kappa score, PPV, AUPRC, and AUROC to ensure a comprehensive and thorough evaluation.

*1) Accuracy*
Accuracy is a performance metric that measures the proportion of correct predictions made by a classification model.

$$Accuracy = (TP + TN)/(TP + FP + TN + FN)$$

Where "TP" stands for the number of true positive samples in a category, while "FN" represents the number of false negative samples. Similarly, "TN" denotes the number of true negative samples, and "FP" indicates the number of false positive samples in a category.

*2) Sensitivity*
The sensitivity indicates the ability of a test to detect positive cases correctly, and a higher sensitivity value indicates a lower rate of false negatives.

$$Sensitivity = TP / (TP + FN)$$

*3) PPV (Positive Predictive Value)*
also known as precision, It quantifies the proportion of correctly predicted positive instances (true positives) out of the total instances predicted as positive (true positives and false positives).

$$PPV = TP / (TP + FP)$$

*4) Kappa score*
kappa is a statistic that measures the agreement between the observed agreement and the expected agreement beyond chance in a classification problem. It is particularly useful when assessing inter-rater reliability.

$$Kappa = (Observed\ agreement - Expected\ agreement) / (1 - Expected\ agreement)$$

The kappa score has a scale of -1 to 1. When it's closer to 1, it means that there is a high level of agreement beyond chance.

*5) AUPRC (Area Under the Precision-Recall Curve):*
It evaluates the trade-off between precision and recall across different classification thresholds, which is helpful for measuring performance in imbalanced datasets. The AUPRC is calculated by computing the area under the precision-recall curve, which plots precision on the y-axis and recall on the x-axis. A higher AUPRC value indicates better model performance in terms of precision and recall trade-off.

| Model | Image size | Sensitivity | Accuracy (%) | kappa | PPV (%) | AUROC | AUPRC |
|---|---|---|---|---|---|---|---|
| MEDUSA [51] | 480x480 | 0.975 | 98.3 | - | 99.0 | - | - |
| ViT-S + SS-CXR [35] | 256x256 | | 98.25 | - | - | 0.999 | 0.999 |
| ViT-Base | 384x384 | 0.98 | 0.99 | 0.98 | 1.0 | 0.9994 | 0.9994 |
| Swin-T-Base | 384x384 | 0.975 | 0.9875 | 0.975 | 1.0 | 0.9988 | 0.9988 |
| MaxVIT | 384x384 | **0.99** | **0.995** | **0.99** | 1.0 | **9.9997** | **0.9997** |
| pvt_v2_b5 | 384x384 | 0.98 | 0.99 | 0.98 | 1.0 | **0.9999** | **0.9999** |

Table. 1 Covid vs. Normal classification Accuracy, kappa score, positive predictive value (PPV), AUROC, and AUPRC of the models on the test samples of the CXR-3 benchmark dataset and comparison to other previous works on this dataset.

*6) AUROC (Area Under the Receiver Operating Characteristic Curve):*

AUROC is another widely used performance metric for binary classification problems. It evaluates the trade-off between true positive rate (sensitivity) and false positive rate (1 - specificity) across different classification thresholds. The AUROC is calculated by computing the area under the ROC curve, which plots the true positive rate on the y-axis against the false positive rate on the x-axis. The AUROC provides a single value to assess the overall performance of a classification model, with higher values indicating better discrimination between positive and negative cases.

*D. Experimental Settings*

We trained all the models using transfer learning, where the transformer networks are initialized with pre-trained weights on ImageNet [48]. We train models on the training set and report top-1 error on the test set. We use Adam optimizer [49] with a momentum of 0.9, a mini-batch size of 8, and a weight decay $5 \times 10^{-2}$ by default. The learning rate is initialized to $5 \times 10^{-5}$ and decreases with cosine schedule [50]. All models are trained to 100 epochs using cross-entropy loss, with warmup epochs of 5. During training, data augmentation techniques like random cropping, horizontal flip, vertical flip, and color jittering have been applied randomly to the training set. Pytorch-based models from Timm [51] have been used for experiments and trained on RTX 3090. The best model during training is selected based on accuracy.

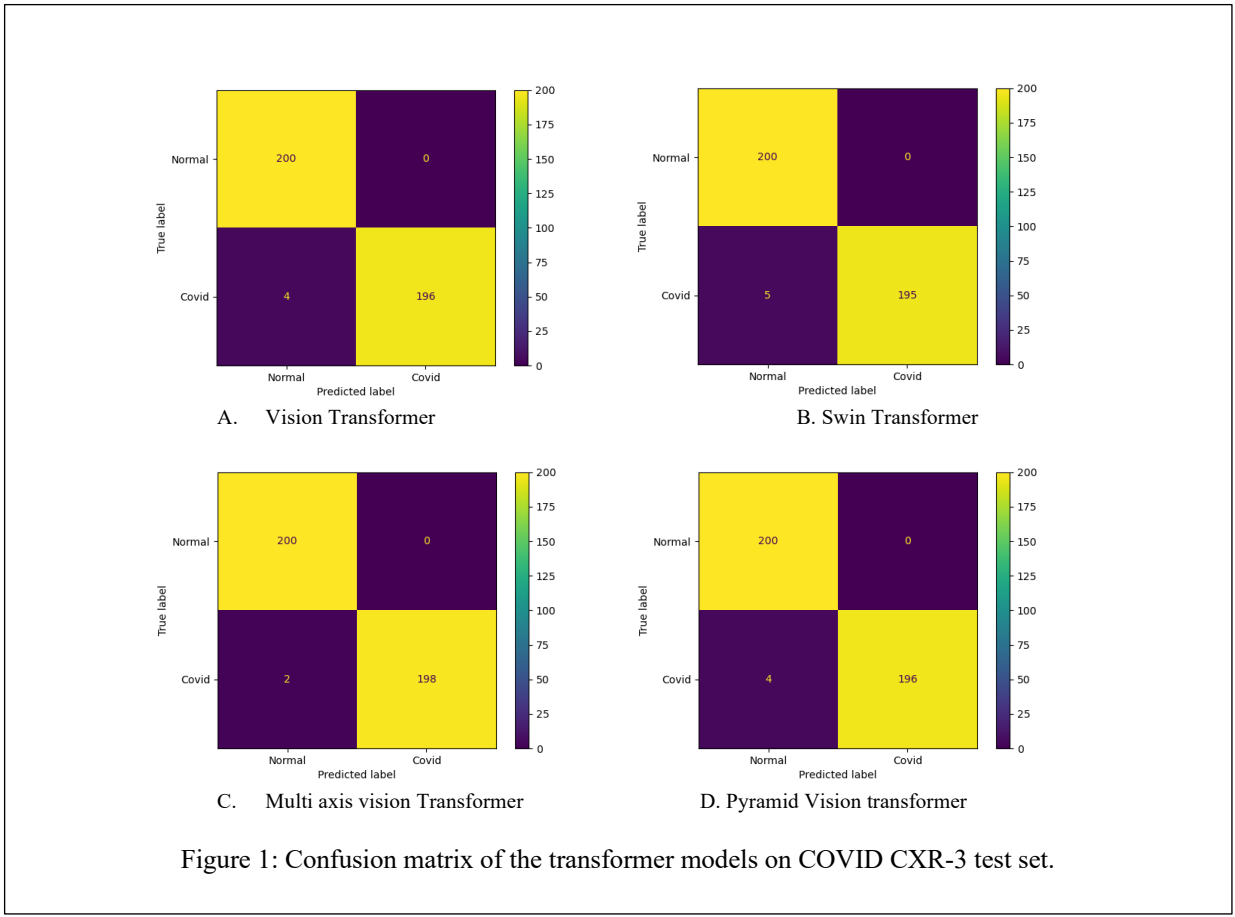

Figure 1: Confusion matrix of the transformer models on COVID CXR-3 test set.

A. Vision Transformer
B. Swin Transformer
C. Multi axis vision Transformer
D. Pyramid Vision transformer

## IV. Results and Discussion

The results of the models suggested for the CXR-3 dataset are presented in Table 1. To evaluate the models, accuracy, kappa score, positive predictive value, AUROC, and AUPRC were used on the test dataset. MaxViT demonstrated the highest accuracy of all the techniques, surpassing the previous best-performing models by 1%. The confusion matrix for the test set, illustrated in Figure 1, shows that MaxViT had fewer misclassifications in comparison to the other models. When it comes to CXRs, MaxViT exhibited slightly superior performance than all the other transformer-based models, making it a highly efficient model for covid classification.

From Table 1, vision transformers with transfer learning clearly achieved state-of-art performance in COVID-19 detection from CXR images. The main reason for Vision Transformers to dominate the existing methods is to utilize self-attention mechanisms, allowing the model to focus on relevant parts of the input image. This helps capture long-range dependencies and relationships between image features, which is particularly beneficial for understanding complex visual patterns and relationships. Vision Transformers capture global context by considering the entire image as a whole. This global perspective gives the model a more holistic understanding of the image and helps make better predictions. Vision Transformers can benefit from transfer learning, where pre-trained models on large-scale datasets can be fine-tuned on specific tasks. This allows the model to leverage knowledge learned from vast amounts of data, leading to improved performance, especially when labeled data is limited.

## V. Conclusion

The use of deep learning techniques in analyzing CXR images holds great potential for COVID-19 diagnosis. Specifically, we have found that transformers-based deep learning algorithms effectively learn long-range dependencies; when used in conjunction with transfer learning and appropriate hyperparameters, these algorithms have demonstrated impressive results. However, we recognize that the study may be limited to this COVIDx CXR-3. Further work involves exploring vision transformers on other medical datasets.